\documentclass{PoS}

\usepackage{epsfig}
\usepackage{graphicx}

\title{Lattice NRQCD study of in-medium bottomonium states using $N_f
  = 2+1, 48^3 \times 12$ HotQCD configurations}

\ShortTitle{In-Medium Bottomonium from NRQCD}

\author{\speaker{Seyong Kim} \\
        Department of Physics, Sejong University, Seoul 143-747, Korea \\
        E-mail: \email{skim@sejong.ac.kr}}

\author{Peter Petreczky \\
        Physics Department, Brookhaven National Laboratory, Upton,
        NY11973, USA \\
        E-mail: \email{petreczk@quark.phy.bnl.gov}}

\author{Alexander Rothkopf \\
        Albert Einstein Center, University of Bern, Sidlerstr. 5, CH-3012 Bern, Switzerland\\
        E-mail: \email{rothkopf@itp.unibe.ch}}

\abstract{
  The behavior of bottomonium state correlators at non-zero
  temperature, $140.4 (\beta = 6.664) \le T \le 221 (\beta = 7.280)$
  (MeV), where the transition temperature is $154(9)$ (MeV), is
  studied, using lattice NRQCD on $48^3 \times 12$ HotQCD HiSQ action
  configurations with light dynamical $N_f = 2 + 1$ ($m_{u,s}/m_s = 0.05$)
  staggered quarks. In order to understand finite temperature effects
  on quarkonium states, zero temperature behavior of bottomonium
  correlators is compared based on $32^4$ ($\beta = 6.664, 6.800$ and
  $6.950$) and $48^3 \times 64$ ($\beta = 7.280)$ lattices.  We find
  that temperature effects on S-wave bottomonium states are small but
  P-wave bottomonium states show a noticeable temperature dependence
  above the transition temperature.
}

\FullConference{The 31th International Symposium on Lattice Field Theory - Lattice 2013,\\
		July 29-August 3, 2013\\
		Mainz, Germany}

\begin{document}
% 7 pages

\vspace{-0.3cm}
\section{Introduction}

Studying the properties of the Quark-Gluon Plasma (QGP) requires us to
compare high energy hadron- and nuclear collisions. In understanding
the differences between the two, we can shed light on the question,
how the presence of a hot medium modifies the abundance of measured
particles. Quarkonium physics offers an excellent opportunity for such
a comparison \cite{Brambilla:2010cs} because inclusive productions and
decays of these quark anti-quark pairs are better understood than
those involving light hadrons. The heavy quark mass of the quarkonium
constituents provides a ``factorization'' of the short distance
perturbative physics from the long distance non-perturbative effects,
which can be discussed by applying various effective field theories of
QCD \cite{Bodwin:1994jh,Brambilla:2004jw}.

\begin{figure}[t]
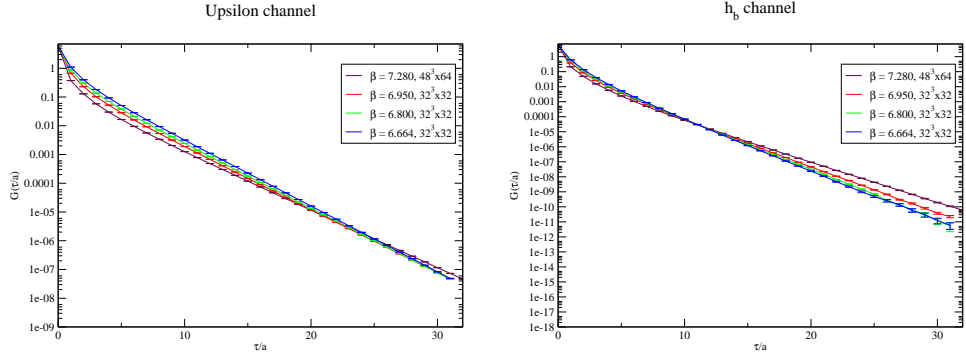

\begin{center}
\epsfig{figure=3S1corr0T.eps,width=0.4\textwidth}
\hspace{0.5cm}\epsfig{figure=1P1corr0T.eps,width=0.4\textwidth}
\end{center}
\caption{S-wave ($\Upsilon$) channel (left) and P-wave ($h_b$) channel
  (right) correlator at $T\simeq0$ temperature for $\beta = 6.664, 6.800,
  6.950,$ and $7.280$ from lattice NRQCD correlator}
 \label{fig:1}
\end{figure}

Recent studies of in-medium quarkonium behavior, based on lattice
NRQCD at finite temperature, show a survival of S-wave bottomonium up
to $\sim 2 T_c$ and melting of P-wave bottomonium above the transition
temperature ($T_c$) \cite{Aarts:2010ek,Aarts:2013kaa}, sequential
suppression of excited states S-wave bottomonium above $T_c$
\cite{Aarts:2011sm}, heavy quark mass dependence of S-wave bottomonium
\cite{Kim:2012by} and velocity dependence of S-wave bottomonium moving
in a thermal environment \cite{Kim:2012by,Aarts:2012ka}. In these
studies, temperature is changed by varying the number of
time-directional lattice slices at a fixed lattice spacing ($T =
\frac{1}{N_\tau a_\tau}$. i.e. $a_\tau$ is fixed and $N_\tau$ is an
integer (= 16, 18, 20, 24, 28, 32, and 80)). Due to the common
renormalization scale this makes it easier to compare the temperature
dependent behavior of lattice NRQCD bottomonium correlators and their
spectral functions at different temperatures.  On the other hand, the
discreteness of the accessible temperatures hampers a detailed study
near $T_c$. In addition, in these studies, the spectral functions are
computed using the standard Maximum Entropy Method (MEM)
\cite{Asakawa:2000tr} on NRQCD bottomonium correlators with NRQCD
kernel ($K(\tau,\omega) = e^{-\omega \tau}$). It will be interesting
to calculate the spectral functions with methods other than the
MEM. Here, we report preliminary results from our lattice NRQCD study
of in-medium bottomonium behavior on $N_f = 2 + 1$ ($m_{u,d}/m_s =
0.05)$ HotQCD configurations, generated with HiSQ fermions on $48^3
\times 12$ \cite{Bazavov:2011nk}. Unlike previous lattice NRQCD
studies \cite{Aarts:2010ek,Aarts:2011sm,Kim:2012by,Aarts:2012ka},
temperature is varied continuously ($N_\tau = 12 = $fixed, $a_\tau$ is
changed by varying the coupling constant) and a novel improved
Bayesian method is used for the computation of spectral functions
\cite{Burnier:2013nla,Burnier:2013fca,Burnier:2013esa}.

\section{Zero Temperature}

Since the lattice spacing in our study is changed to vary temperature,
a comparison of NRQCD correlators at different temperatures is more
complicated. It requires $T\simeq0$ simulations, which supplement the
finite temperature simulations at the same gauge coupling constant. We
determine the quarkonium mass from an exponential fit to the NRQCD
correlators, i.e. an ``energy offset'' needs to be obtained
\cite{Dowdall:2011wh}. Also the performance of the improved Bayesian
method used in this work can be assessed using the zero temperature
features of bottomonium spectral functions. Thus, we first carry out a
lattice NRQCD analysis on low temperature HotQCD configurations. The
list of parameters for these lattice configurations is given in
Tab. (\ref{tab:parameterT0}), where $M_b a$ is chosen for $M_b^{\rm
  exp} = 4.65$ (GeV) for a given $a$.

\begin{table}[ht]
\begin{center}
\vspace*{0.2cm}
\begin{tabular}{|c|c|c|r|r|c|}
\hline
$\beta$& Volume        &$T$(MeV)   & $a$(fm)&$M_b a$ & analyzed Ncfg\\
\hline
6.664  &$32^3\times 32$&52.65 & 0.1169 & 2.759 & 100 \\
6.800  &$32^3\times 32$&59.93 & 0.1027 & 2.424 & 100 \\
6.950  &$32^3\times 32$&68.98 & 0.08925& 2.107 & 100 \\
7.280  &$48^3\times 64$&46.62 & 0.06603& 1.559 & 100 \\
\hline
\end{tabular}
\vspace*{0.2cm}
\caption{Parameter list for the $T\simeq0$ configurations}
\label{tab:parameterT0}
\end{center}
\end{table}

Fig.~(\ref{fig:1}) shows the typical behavior of S-wave and P-wave
bottomonium correlators, which are computed from NRQCD bottom quark
propagators at $T \simeq 0$, as a function of the Euclidean time
($\tau$). Exponential function fits to the quarkonium correlators
yield the energy of the corresponding quarkonium states (e.g., $E_\Upsilon
(1S)$ for the Upsilon channel and $E_{h_b} (1P)$ for the $h_b$
channel, respectively). Here, from ${M_\Upsilon}^{\rm exp} = E_G +
E_{\Upsilon (1S)}$ at each $\beta$, we determine the constant energy
offset for the simulated $\beta$.

\begin{figure}[ht]
\begin{center}
\epsfig{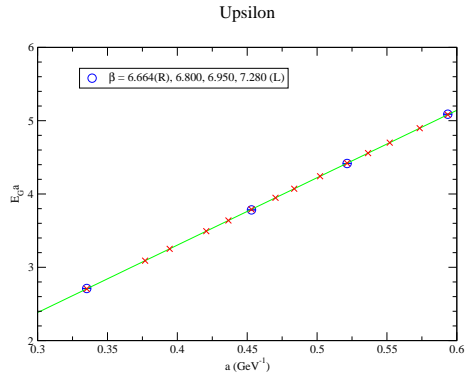}
\end{center}
\caption{``Energy offset constant'' for $\beta = 6.664, 6.800, 6.950,$ and
$7.280$ from 1-exponential fit to $\Upsilon$ correlator (blue circle)}
 \label{fig:2}
\end{figure}

\begin{figure}[ht]
\begin{center}
\epsfig{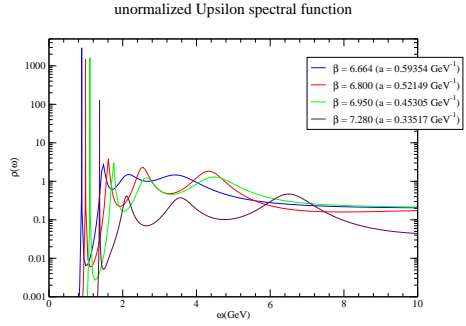}
\end{center}
\caption{Un-normalized $\Upsilon$ spectral functions from the improved
  Bayesian method for $\beta = 6.664, 6.800, 6.950,$ and $7.280$}
 \label{fig:3}
\end{figure}

Fig.~(\ref{fig:2}) shows the energy offset constant ($E_G a$) at each
$\beta$. The blue circles in the figure are the energy offset
determined from the fitted energy of the zero temperature $\Upsilon
(1S)$ state at $\beta = 6.664, 6.800, 6.950, 7.280$ (the horizontal
axis is the lattice spacing, $a^{-1}$ (GeV) for each $\beta$). The
green line is the linear interpolation among the blue circle data
points. The red crosses are the energy offset which will be used for
non-zero temperature runs. In addition, the computed $\Upsilon$
spectral functions for $\beta = 6.664, 6.800, 6.950, 7.280$ at
$T\simeq0$ temperature are shown in Fig.~(\ref{fig:3}). The position
of the first peak agrees with the result of a 1-exponential function
fit to the respective $\Upsilon$ correlators within errors. The shapes
of the spectral functions are quite similar to each other as expected
and may be matched to each other by accommodating the constant energy
offset effect (constant amount shift along the horizontal axis for the
spectral functions) and by considering the normalization difference in
the spectral function between $32^4$ lattices and $48^3 \times 64$
lattice (rescale along the vertical axis for the spectral functions).

\section{Around the Transition Temperature}

\begin{table}[ht]
\begin{center}
\vspace*{0.2cm}
\begin{tabular}{|c|c|c|r|r|c|}
\hline
$\beta$	& T    & $T/T_c$& $a$(fm) &$M_b a$ & analyzed Ncfg\\
\hline
6.664	&140.40& 0.911  & 0.1169  & 2.759  & 100 \\
6.700	&145.32& 0.944  & 0.1130  & 2.667  & 100 \\
6.740	&150.97& 0.980  & 0.1087  & 2.566  & 100 \\
6.770	&155.33& 1.008  & 0.1057  & 2.495  & 100 \\
6.800	&159.80& 1.038  & 0.1027  & 2.424  & 100 \\
6.840	&165.95& 1.078  & 0.09893 & 2.335  & 100 \\
6.880	&172.30& 1.119  & 0.09528 & 2.249  & 100 \\
6.910	&177.21& 1.151  & 0.09264 & 2.187  & 100 \\
6.950	&183.94& 1.194  & 0.08925 & 2.107  & 100 \\
6.990	&190.89& 1.240  & 0.086	  & 2.030  & 100 \\
7.030	&198.08& 1.286  & 0.08288 & 1.956  & 100 \\
7.100	&211.23& 1.371  & 0.07772 & 1.835  & 100 \\
7.150	&221.08& 1.436  & 0.07426 & 1.753  & 100 \\
7.280	&248.63& 1.614  & 0.06603 & 1.559  & 100 \\
\hline
\end{tabular}
\vspace*{0.2cm}
\caption{Parameter list for the $T \neq 0$ configurations}
\label{tab:parameters}
\end{center}
\end{table}

\begin{figure}[ht]
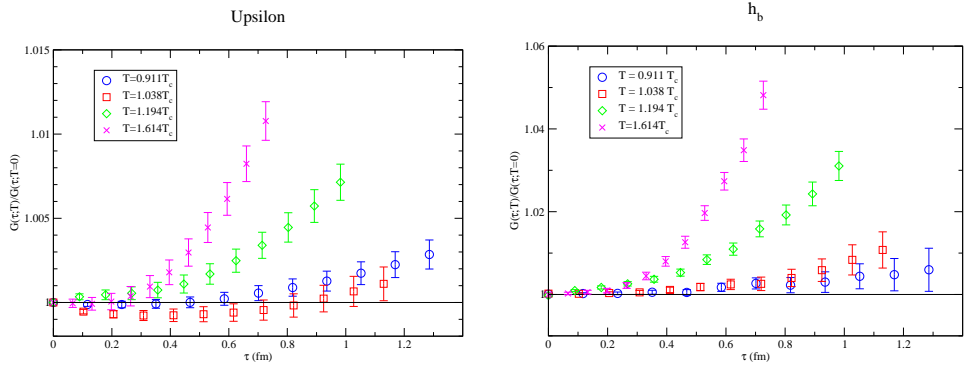

\begin{center}
\vspace{-0.2cm}
\epsfig{figure=3S1ratioN.eps,width=0.4\textwidth}
\hspace{0.5cm}\epsfig{figure=1P1ratioN.eps,width=0.4\textwidth}
\end{center}
\caption{The ratio of $\Upsilon$ channel correlators at non-zero
  temperature to the corresponding correlators at $T\simeq 0$ temperature (left)
  and that of $h_b$ (right) at $\beta = 6.664, 6.800, 6.950,$ and
  $7.280$}
 \label{fig:4}
\end{figure}
\begin{figure}[ht]
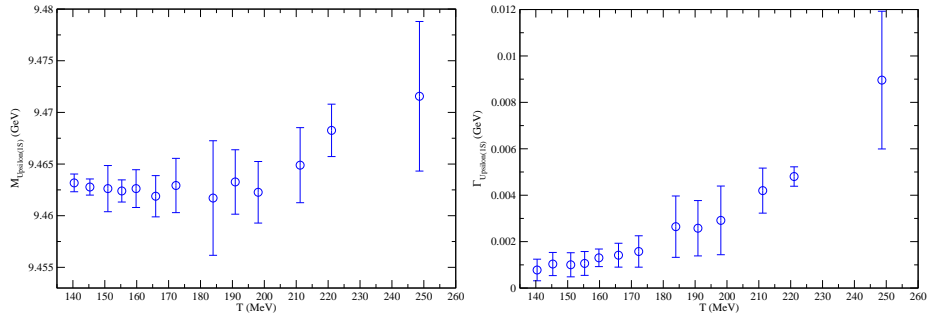

\begin{center}
\vspace{-0.4cm}
\includegraphics[width=0.4\textwidth ,clip=true, trim= 0cm 0cm 0cm 0.08cm]{Swave_Peak_vs_T_new.eps}
\includegraphics[width=0.4\textwidth ,clip=true, trim= 0cm 0cm 0cm 0.08cm]{Swave_Width_vs_T_new.eps}
\end{center}\vspace{-0.2cm}
\caption{$\Upsilon : $ the first peak (1S) position of $\Upsilon$
  channel spectral function vs. temperature (left) and the width of
  the first peak of $\Upsilon$ channel spectral function
  vs. temperature}
 \label{fig:6}
\end{figure}
\begin{figure}[ht]
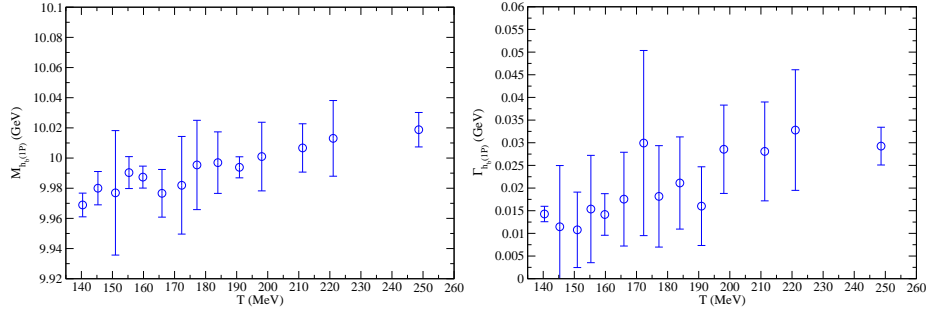

\begin{center}
\includegraphics[width=0.4\textwidth ,clip=true, trim= 0cm 0cm 0cm 0.08cm]{Pwave_Peak_vs_T_new.eps}
\includegraphics[width=0.4\textwidth ,clip=true, trim= 0cm 0cm 0cm 0.08cm]{Pwave_Width_vs_T_new.eps}
\end{center}\vspace{-0.4cm}
\caption{$h_b $: the first peak (1P) position of $h_b$ channel
  spectral function vs. temperature (left) and the width of the first
  peak of $h_b$ channel spectral function vs. temperature}
 \label{fig:7}
\end{figure}

The list of parameters for the non-zero temperature lattice
configurations used in our lattice NRQCD analysis is given in
Tab. (\ref{tab:parameters}). Fig.~(\ref{fig:4}) shows the ratios of
the non-zero temperature quarkonium correlators ($48^3 \times 12$) to
the $T\simeq0$ temperature quarkonium correlators ($32^4$ and $48^3
\times 64$) at those values of $\beta$ at which $T\simeq0$
configurations are available ($\Upsilon$ and $h_b$ channel). In these
ratios, the energy offset effect cancels, since it is a function of
lattice spacing and is independent of the temperature. Similar to the
results from previous studies
\cite{Aarts:2010ek,Aarts:2011sm,Kim:2012by,Aarts:2012ka}, the S-wave
ratios exhibit only small changes with temperature, while the P-wave
ratios show a large $T$ dependence. In contrast, the advantage of
continuously changing temperature becomes clear, when we compare the
ratios from all the temperatures. The temperature effect in the P-wave
increases as $T$ surpasses $T_c$. Interestingly however the
temperature effect in the S-wave at $T = 0.911 T_c$ is more or less
similar to that at $T = 1.038 T_c$ begins to increase only at $T \ge
1.194 T_c$.
\begin{figure}[ht]
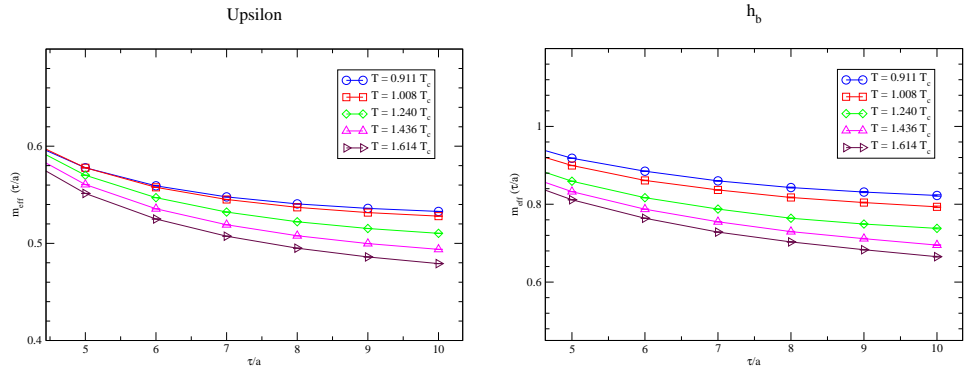

\begin{center}
\epsfig{figure=effmass3S1sub.eps,width=0.4\textwidth}
\hspace{0.5cm}\epsfig{figure=effmass1P1sub.eps,width=0.4\textwidth}
\end{center}
\vspace{-0.4cm}
\caption{Effective mass for the $\Upsilon$ channel correlators (left) and
  that for $h_b$}
 \label{fig:5}
\end{figure}
In Fig.~(\ref{fig:5}) we plot $m_{\rm eff} (\tau) = - \ln \left[ G
  (\tau + a) / G (\tau) \right]$. Instead of showing the effective
mass data from all 14 temperatures in Tab. (\ref{tab:parameters}), we
chose five temperatures for a clear presentation.  None of the
effective mass plots contain a clear plateau. Similar to
Fig.~(\ref{fig:4}), the S-wave effective mass plot exhibits a small
effect below $T_c$ and shows an increasing temperature dependence
above $T_c$. In contrast, the P-wave effective mass plot shows an
decrease in the values for the whole temperature range. We also
applied the new Bayesian method to non-zero temperature bottomonium
correlators to obtain the spectral functions for each channel. In
Fig.~(\ref{fig:6}), we show the first peak position of the S-wave
($\Upsilon$) channel spectral function and the width of the first peak
at each temperature.  Fig.~(\ref{fig:7}) contains the values for the
P-wave ($h_b$). The error bars in each figure are estimated from the
Jacknife spread of measured values (peak position and its width) on 5
sets of 20 correlators for each channels. Similar to the observations
from Fig.~(\ref{fig:4},\ref{fig:5}), the S-wave spectral functions
behave differently from the P-wave spectral functions. The first peaks
of the S-wave spectral functions stay at the same position below $T_c$
and begin to move upward above $T_c$ as $T$ increases. The position of
the first peak of the P-wave spectral functions appears to rise
linearly in $T$ across $T_c$. Also, the width of the first peak of the
P-wave spectral functions seems to have a $T$-dependence albeit large
Jackknife error bar.

\section{Conclusion}

We have investigated the non-zero temperature behavior of S-wave and
P-wave bottomonium by computing lattice NRQCD correlators on $N_f =
2+1 (m_{u,d}/m_s = 0.05)$ flavor HotQCD configurations and analyzing
their spectral functions using an improved Bayesian method
\cite{Burnier:2013nla,Burnier:2013fca,Burnier:2013esa}. The studied
temperature is $140.4$ (MeV) $\le T \le 221$ (MeV), where the
cross-over transition temperature is $154(9)$ MeV. Compared to the
previous lattice NRQCD studies on quarkonium in-media, where the
lattice spacing is fixed and the temperature can only be changed
discretely by changing the number of the time slices
\cite{Aarts:2010ek,Aarts:2011sm,Kim:2012by,Aarts:2012ka}, temperature
can be varied continuously in this study. Hence the modification of
quarkonium across the transition temperature can be investigated in
detail.

Preliminary results show that the behavior of S-wave bottomonium
(Upsilon and $\eta_b$ states) is distinctly different from that of
P-wave bottomonium ($h_b$ state) across $T_c$. The ratio of the $T>0$
correlator to the $T\simeq0$ temperature correlator, as well as the
effective mass for the S-wave is more or less unchanged below $T_c$
and begins to increase only above $T_c$. The temperature effect in the
ratio and in the effective mass for the P-wave appears to be $\propto
T$. Spectral functions from the improved Bayesian method show a very
similar picture. The first peak position of S-wave bottomonium
spectral functions stays more or less the same below $T_c$ and begins
to increase above $T_c$. On the other hand, the first peak position of
P-wave bottomonium spectral functions appears to move proportional to
$T$ across $T_c$. Further analysis with increased statistics and
systematic error analysis of the improved Bayesian method will follow
in near future.

{\bf Acknowledgements}

SK is supported by the National Research Foundation of Korea grant
funded by the Korean government (MEST) No.\ 2012R1A1A2A04668255. PP is
supported by U.S.Department of Energy under Contract
No.DE-AC02-98CH10886. AR is partly supported by the Swiss National
Science Foundation (SNF) under grant 200021-140234.

\end{document}